\documentclass[a4paper]{article}

\usepackage{amsmath}
\usepackage{amsfonts}
\usepackage{dsfont}

\newcommand{\absb}[1]{|#1|}

\newcommand{\m}[1]{$#1$}

\newcommand{\refeqp}[1]{(\ref{eq:#1})}

\newcommand{\paraa}[1]{\big(#1\big)}
\newcommand{\parab}[1]{\Big(#1\Big)}
\newcommand{\parac}[1]{\bigg(#1\bigg)}

\newcommand{\real}{\mathbb{R}}

\newcommand{\half}{\frac{1}{2}}
\newcommand{\third}{\frac{1}{3}}

\newcommand{\bin}[2]{\begin{pmatrix} #1 \\ #2 \end{pmatrix}}
\renewcommand{\mid}{\mathds{1}}

\newcommand{\Mv}{{\overrightarrow{\! M}}}

\newcommand{\A}{\mathcal{A}}
\newcommand{\su}{\textrm{su}}

\newcommand{\X}{X}
\newcommand{\R}{R}
\newcommand{\Xdd}{\overset{\, ..}{\X}}
\newcommand{\Xd}{\overset{\, .}{\X}}

\newcommand{\xd}{\overset{.}{x}}
\newcommand{\const}{\textrm{const}}
\newcommand{\CCom}[3]{\Big[\big[#1,#2\big],#3\Big]}

\newcommand{\Eab}{E_{ab}}
\newcommand{\Eba}{E_{ba}}
\newcommand{\Ebc}{E_{bc}}

\newcommand{\Epmab}{E^{\pm}_{ab}}

\newcommand{\Epbc}{E^+_{bc}}

\newcommand{\Embc}{E^-_{bc}}
\newcommand{\Eaa}{E_{aa}}

\newcommand{\eabc}{\epsilon_{abc}}
\newcommand{\Ma}{M_{a}}
\newcommand{\Map}{M_{a'}}
\newcommand{\Mapp}{M_{a''}}
\newcommand{\Mha}{\hat{M}_{a}}
\newcommand{\Mhap}{\hat{M}_{a'}}
\newcommand{\Mhapp}{\hat{M}_{a''}}

\newcommand{\Dm}{\Delta_{-}}
\newcommand{\Dp}{\Delta_{+}}
\newcommand{\Dpp}{\Delta_{||}}
\newcommand{\diag}{\operatorname{diag}}
\newcommand{\diagmatrix}[3]{\begin{pmatrix} #1&0&0 \\ 0&#2&0 \\ 0&0&#3\end{pmatrix}}
\newcommand{\twomatrix}[4]{\begin{pmatrix} #1 & #2 \\ #3 & #4 \end{pmatrix}}
\newcommand{\SO}{\textrm{SO}}
\newcommand{\id}{\operatorname{id}}
\renewcommand{\R}{\mathcal{R}}

\begin{document}

\thispagestyle{empty}

\vspace{2cm}

\begin{center}
\Large{\textbf{Classical Solutions in the BMN Matrix Model}}\\
\vspace{0.25cm}
\large{Joakim Arnlind and Jens Hoppe}\\
\small{Department of Mathematics\\
Royal Institute of Technology\\
Stockholm\\
\vspace{0.3cm}
 December 2003}

\vspace{3cm}

\textbf{\large{Abstract}}\\
\end{center}


\noindent Several reductions of the bosonic BMN matrix model equations
to ordinary point particle Hamiltonian dynamics in the plane (or
\m{\real^3}) are given -- as well as a few explicit solutions (some of
which, as \m{N\rightarrow\infty}, correspond to membranes rotating
with constant angular velocity, others to higher dimensional objects).

\newpage

\pagenumbering{arabic}

\noindent Consider the following 9 hermitean \m{3\times 3} matrices
(\m{a=1,2,3;a'=a+3;a''=a+6}, \m{\paraa{E_{ab}}_{cd}=\delta_{ac}\delta_{bd}})
\begin{align}
  \begin{split}
    \Mha &= -i\eabc\Ebc\\
    \Mhap &= \Eaa-\third\mid\\
    \Mhapp &= \absb{\eabc}\Ebc,
  \end{split}\label{eq:Madef}
\end{align}
\m{\parab{\sum \Mha^2=3\sum\Mhap^2=\sum\Mhapp^2=2\cdot\mid}}\\
\noindent and the corresponding discrete Laplace-operators
\begin{align}
  \begin{split}
    \Dm &:= \CCom{\,\cdot\,}{\Mha}{\Mha} =
    -\half\CCom{\,\cdot\,}{\Embc}{\Embc}\\
    \Dpp &:= \CCom{\,\cdot\,}{\Mhap}{\Mhap} =
    \CCom{\,\cdot\,}{\Eaa}{\Eaa}\\
    \Dp &:= \CCom{\,\cdot\,}{\Mhapp}{\Mhapp} =
    \sum_{b<c}\CCom{\,\cdot\,}{\Epbc}{\Epbc},\\
  \end{split}\label{eq:Laplacedef}
\end{align}
where \m{\Epmab:=\Eab\pm\Eba}. As is easy to check, the action of
\refeqp{Laplacedef} on \refeqp{Madef} is purely diagonal, with
eigenvalues given according to 
\begin{align}
  \begin{split}
    \Dm &= \diag \paraa{222\, 666\, 666}\\
    \Dpp &= \diag \paraa{222\, 000\, 222}\\
    \Dp &= \diag \paraa{666\, 666\, 222},
  \end{split}\label{eq:diagaction}
\end{align}
and because \refeqp{Laplacedef} only involves commutators, \refeqp{diagaction} extends
to the action of \refeqp{Laplacedef} on the 9 \m{N\times N} matrices
corresponding to \refeqp{Madef} in an arbitrary \m{N}-dimensional
representation! of \m{\su(3)} (just write \refeqp{Madef} in terms of
the Cartan Weyl basis of \m{\su(3)}).

This way one can find 9 hermitean \m{N\times N} matrices
\m{M_{i=1,\ldots,9}}, resp. a matrix valued 9-vector \m{\Mv}
satisfying
\begin{align}
  \begin{split}
    \CCom{\Mv}{\Ma}{\Ma} &= 2\diagmatrix{\mid}{3\mid}{3\mid}\Mv\\
    \CCom{\Mv}{\Map}{\Map} &= 2\diagmatrix{\mid}{0}{\mid}\Mv\\
    \CCom{\Mv}{\Mapp}{\Mapp} &= 2\diagmatrix{3\mid}{3\mid}{\mid}\Mv.
  \end{split}\label{eq:Mveq}
\end{align}
We will now use this to find nontrivial, timedependent solutions of
the bosonic BMN \cite{BMN} matrix model equations, 
\begin{align}
  \begin{split}
    \Xdd_a &= - \sum_{i=1}^9\CCom{X_a}{X_i}{X_i} -
    4m^2X_a -3im\eabc[X_b,X_c]\\
    \Xdd_\mu &= -\sum_{i=1}^9 \CCom{X_\mu}{X_i}{X_i}
    - m^2X_\mu\quad(\mu=4,5,\ldots,9)\\
    &\sum_{i=1}^9 [X_i,\Xd_i]=0.
  \end{split}\label{eq:Xcom}
\end{align}
Letting, e.g. , 
\begin{align}
  \begin{split}
    X_a(t) &= x(t)\Ma,\,\, X_{a'}(t) = \sqrt{3}\,y(t)\Map,\,\,
    X_{a''}(t) = z(t)\Mapp,
  \end{split}\label{Xadef}
\end{align}
with \m{\Mv=\parab{(\Ma)(\Map)(\Mapp)}} satisfying \refeqp{Mveq} (and
\m{[\Ma,M_b]=i\eabc M_c}) reduces \refeqp{Xcom}, for arbitrary \m{N}, to differential
equations involving only 3 scalar functions (\m{x,y} and \m{z}):
\begin{align}
  \begin{split}
    & \overset{\,..}{x} + x\paraa{4m^2+2x^2+6y^2+6z^2-6mx}=0\\
    & \overset{\,..}{y} + y\paraa{m^2+6x^2+6z^2}=0\\
    & \overset{\,..}{z} + z\paraa{m^2+6x^2+6y^2+2z^2}=0,
  \end{split}\label{eq:xyzdiff}
\end{align}
which are, in a canonical way, Hamiltonian with respect to
\begin{align}
  \begin{split}
    \half\parab{\dot{x}^2+\dot{y}^2+\dot{z}^2}&
    +3\parab{x^2y^2+x^2z^2+y^2z^2}\\
    &+\half m^2\parab{4x^2+y^2+z^2}
    +\half z^4 + \half x^4 -2mx^3.
  \end{split}\label{eq:hamiltonian}
\end{align}
Another reduction can be obtained by letting (cp. \cite{AH})
\begin{align}
  \begin{split}
    X_a(t) &= x(t)M_a,\,\, 
    X_\mu = \sqrt{\frac{3}{5}}z(t)\R_{\mu\nu}(t)\tilde{M}_{\nu}\\
    \parab{R_{\mu\nu}} &= e^{\mathcal{A}\varphi(t)}\in\SO(6),\,\,
    z^2\dot{\varphi}(t) = L = \const,\,\,
    \mathcal{A}^2 = -\id;\\
    \tilde{M}_{a'} &:= \sqrt{2}\Map,\,\, \tilde{M}_{a''}:=\Mapp
  \end{split}\label{eq:XReq}
\end{align}
(note that \m{[\tilde{M}_{a'},\tilde{M}_{a''}] = 0}), yielding
\begin{align}
  \begin{split}
    &\overset{\,..}{x} + x\paraa{4m^2+2x^2+6z^2-6mx}=0\\
    &\overset{\,..}{z} + z\paraa{m^2+6x^2+\frac{18}{5}z^2-\frac{L^2}{z^4}}=0.
  \end{split}\label{eq:zxdiff}
\end{align}
For the Ansatz \refeqp{XReq} to work it is important
that all 6 \m{N\times N} matrices \m{\tilde{M}_\nu} have the same eigenvalue under
the action of both \m{\Dm} and \m{\tilde{\Delta}_+ +2\tilde{\Delta}_{||}}.

Various other choices and combinations are possible, e.g.
\m{\tilde{M}_{a''}=0}, \m{\tilde{M}_{a'}=M_a}, i.e.
\begin{align}
  \begin{split}
    X_a &= x(t)M_a\\
    X_\mu &= z(t)
    \parab{\cos\varphi M_1,\cos\varphi M_2,\cos\varphi M_3,
    \sin\varphi M_1,\sin\varphi M_2,\sin\varphi M_3}\\
    z^2\dot{\varphi} &= L = \const,
  \end{split}\label{eq:Xfuzzy}
\end{align}
giving
\begin{align}
  \begin{split}
    &\overset{\,..}{x} + x\paraa{2x^2+2z^2+4m^2-6mx} = 0\\
    &\overset{\,..}{z} + z\paraa{2x^2+2z^2+m^2-\frac{L^2}{z^4}} = 0.
  \end{split}\label{eq:xzsymdiff}
\end{align}
Apart from the trivial static (known) solutions, (\m{L=0,z=0;x=0,m} or
\m{2m}), and genuinely timedependent solutions of \refeqp{xzsymdiff},
there are several ``intermediate'' solutions, for which \m{z} is
constant, but non-zero (making \m{\varphi(t)} linear in \m{t}): 2 for
which \m{x=0,z=\pm z_0}, as well as those corresponding to the roots
of the quintic equation obtained via \m{z^2=3mx-x^2-2m^2}.

Replacing \m{M_a} by \m{\Map}, resp. \m{\Mapp}, in the second part
of \refeqp{Xfuzzy} just changes the
\m{\twomatrix{2}{2}{2}{2}\bin{x^2}{z^2}} part in \refeqp{xzsymdiff} to 
\begin{align}
  \twomatrix{2}{2}{6}{0}\bin{x^2}{z^2} \textrm{ resp. } \twomatrix{2}{6}{6}{2}\bin{x^2}{z^2},
\end{align}
leading to yet other solutions.

Of course one can consider both the \m{m\rightarrow 0}
(\m{m\rightarrow\infty}) limit of these solutions as well as their
\m{N\rightarrow\infty} continuum limit.

Finally note that one can also let both \m{X_\mu} \emph{and} \m{X_a} rotate,
letting e.g.
\begin{align}
  \begin{split}
    X_a(t) &= \sqrt{6}\,x(t)
    \paraa{\cos\theta M_4 - \sin\theta M_5,\sin\theta M_4 + \cos\theta
    M_5,M_6}\\
    X_\mu(t) &= y(t)\R_{\mu\nu}\tilde{M}_\nu\\
    \tilde{M}_{a'} &= \Ma,\,\, \tilde{M}_{a''} = \Mapp\\
    x^2\dot{\theta} &= K,\,\, y^2\dot{\varphi} = L
  \end{split}
\end{align}
(as before, \m{\R = e^{\A\varphi(t)},\ldots})\\
\noindent which result in equations of motion,
\begin{align}
  \begin{split}
    &\overset{\,..}{x} + x\parac{4m^2+12y^2-\frac{K^2}{x^4}} = 0\\
    &\overset{\,..}{y} + y\parac{m^2+12x^2+8y^2-\frac{L^2}{y^4}} = 0,
  \end{split}\label{eq:xzrotate}
\end{align}
corresponding to
\begin{align}
  \begin{split}
    H = \half\parab{\xd^2+\dot{y}^2} + \frac{L^2}{2y^2} +
    \frac{K^2}{2x^2} + \frac{\,\,m^2}{2}\parab{y^2+4x^2} + 6x^2y^2+2y^4
  \end{split}
\end{align}

\section*{Acknowledgement}

\noindent We thank C. Zachos for correspondence, and J. Plefka for
discussing with us the behaviour under supersymmetry-transformations
of (such) classical solutions and his joint
work with N. Kim (hep-th/0207034) on
protected states.

\bibliographystyle{unsrt}

\end{document}